\documentclass[aps,prb,twocolumn,superscriptaddress,showkeys,citeautoscript]{revtex4-1}

\usepackage[colorlinks=true,bookmarks=false,citecolor=blue,linkcolor=red,hyperfootnotes=true,urlcolor=blue]{hyperref}

\usepackage{physics}
\usepackage{graphicx}
\usepackage{dcolumn}
\usepackage{bm}
\usepackage[dvipsnames]{xcolor}
\usepackage{wasysym}
\usepackage{lipsum}
\usepackage{enumitem}
\usepackage{booktabs} 
\usepackage{gensymb}
\usepackage{xcolor}
\usepackage{amssymb}
\usepackage{booktabs}
\usepackage{array}
\newcolumntype{L}{>{$}l<{$}}
\newcolumntype{C}{>{$}c<{$}}
\newcolumntype{R}{>{$}r<{$}}

\usepackage{blindtext}

\makeatletter
\def\p@subsection{}
\makeatother

\newcommand{\vecb}{\mathbf{b}}

\newcommand{\vecq}{\mathbf{q}}
\newcommand{\vecn}{\mathbf{n}}
\newcommand{\vecf}{\mathbf{f}}

\newcommand{\veck}{\mathbf{k}}

\newcommand{\code}[1]{\texttt{#1}}
\newcommand{\ii}{\code{i}}

\begin{document}

\title{Numerically Exact Generalized Green's Function Cluster
Expansions for Electron-Phonon Problems}

\author{Matthew R. Carbone}\email{mrc2215@columbia.edu}
\affiliation{Department of Chemistry, Columbia University, New York,
New York 10027, USA}

\author{David R. Reichman}
\affiliation{Department of Chemistry, Columbia University, New York,
New York 10027, USA}

\author{John Sous}\email{js5530@columbia.edu}
\affiliation{Department of Physics, Columbia University, New York,
New York 10027, USA}

\date{\today}

\begin{abstract}
We generalize the family of approximate momentum average methods to formulate a numerically exact, convergent hierarchy of equations whose solution provides an efficient algorithm to compute the Green's function of a particle dressed by bosons suitable in the entire parameter regime. We use this approach to extract ground-state properties and spectral functions. Our approximation-free framework, dubbed the generalized Green's function cluster expansion (GGCE), allows access to exact numerical results in the extreme adiabatic limit, where many standard methods struggle or completely fail. We showcase the performance of the method, specializing three important models of charge-boson coupling in solids and molecular complexes: the molecular Holstein model, which describes coupling between charge density and local distortions, the Peierls model, which describes modulation of charge hopping due to intersite distortions, and a more complex Holstein+Peierls system with couplings to two different phonon modes, paradigmatic of charge-lattice interactions in organic crystals. The GGCE serves as an efficient approach that can be systematically extended to different physical scenarios, thus providing a tool to model the frequency dependence of dressed particles in realistic settings.
\end{abstract}


\maketitle

\section{Introduction}
The interaction of a particle with its environment is central to the
study of many physical systems. 
One classic problem of this type is that of the polaron, which describes a 
mobile carrier dressed by bosonic fluctuations.\cite{mahan2013many} Originally
predicted by Landau,\cite{landauOG}
expanded upon by Pekar~\cite{pekar1946local,landau1948effective} and cemented into condensed
matter canon by Lee, Low and Pines,\cite{lee1953motion} Fr{\"o}hlich, Pelzer and
Zienau,\cite{frohlich1950xx,frohlich1954electrons}
Feynman,\cite{feynman1955slow} and Holstein,\cite{holstein1959studiesI,holstein1959studiesII}
a polaron forms when a particle such as  an electron or hole moves in a deformable
medium. The motion of the particle induces a local polarization cloud, which is dragged along with the particle
as it moves, renormalizing its effective mass and yielding a non-zero quasiparticle weight. Polarons arise
in a variety of physical contexts beyond that of electron-phonon systems,\cite{alexandrov1994bipolarons} such as excitons in photoexcited molecular crystals,\cite{scholes2006,park2009bulk,spano2010spectral,XYZ2017Polaron}
hole-doped magnets,\cite{trugman1988interaction} light-matter systems,\cite{Basov2016,Feist2018,Xiang2020} impurities
embedded in ultracold gases\cite{schirotzek2009observation,koschorreck2012attractive,jorgensen2016observation,hu2016bose} and in other more exotic physical settings.\cite{baggioli2015electron,Sous2020Fracton1,Sous2020Fracton2}

Over the last two and a half decades, many (in principle) exact numerical methods have been devised to study polaronic problems. One can broadly classify these approaches into two main categories: real- and imaginary-frequency methods. Approaches in the former class include Variational Exact Diagonalization,\cite{bonvca1999holstein} and its variants,\cite{bonca2,bonca3} Limited Phonon Basis Exact Diagonalization\cite{de2009optical} and Matrix-Product-State techniques\cite{jeckelmann1998density,Jeckelmann2,FHM,BK}. Methods in the latter class are most prominently Monte Carlo methods, such as Diagrammatic,\cite{prokof1998polaron,prokof2008fermi,mishchenko2000diagrammatic}, Path-integral\cite{titantah2001free} and Continuous-time\cite{kornilovitch1998continuous} Monte Carlo. While Monte Carlo techniques are well suited for the study of finite-temperature systems over the complete range of polaronic model parameters, they require ill-conditioned analytic continuation to the real-frequency axis in order to study dynamics.~\footnote{We note that recent advances in Diagrammatic Quantum Monte Carlo enable the summation of graphs on the real-frequency axis, see e.g. Refs.~\onlinecite{taheridehkordi2019algorithmic,taheridehkordi2020optimal,vuvcivcevic2020real}. Such techniques may soon enable the direct calculation of spectral information in the type of models we consider here.} In contrast, direct real-time methods face a daunting challenge in several parameter regimes, including the so-called adiabatic limit where the lattice response is slow, as well as the strong-coupling limit, where a large number of bosons is excited in the system and the size of basis states becomes too large to efficiently manage.

In this work, we introduce the Generalized Green's function Cluster Expansion (GGCE), a non-perturbative approach
that enables an exact, efficient numerical computation of real-frequency Green's functions of polaronic models even in regimes challenging for related real-frequency approaches. 
We restrict ourselves to the limiting case of a single carrier in an otherwise unoccupied band~\cite{mahan2013many,dunn1975electron}, reserving an attempt to formulate a cluster expansion approach for the real-frequency properties of polaron models at finite concentrations~\footnote{Determinant Quantum Monte Carlo and its variations~\cite{zhang2019charge,li2020quantum,li2019enhancement,lee2021constrained} has been shown to perform well for these problems.} for future work.
In particular, we show that the GGCE provides access to exact spectra in the portions of the adiabatic and
strong-coupling limits inaccessible to more standard Variational Exact Diagonalization approaches, while converging more rapidly in accessible regimes. Our method builds on the Momentum Average (MA) Approximation,\footnote{
The MA approach has been validated for a large number of systems,
including, but not limited to, Holstein,\cite{berciu2006green,goodvin2006green,berciu2007systematic,covaci2007holstein,goodvin2009holstein,berciu2010holstein}, Peierls,\cite{marchand2010sharp} Edwards,\cite{berciu2010momentum} and dual-coupled polarons,\cite{marchand2017dual} Holstein\cite{AdolphsBipolaron} and Peierls bipolarons,\cite{SousBipolaron} and has been applied to model experimental systems such as graphene\cite{covaci2008survival} and cuprates,~\cite{ebrahimnejad2014dynamics} for example} proposed by Berciu in 2006,\cite{berciu2006green} which
has since been adapted to describe realistic materials~\cite{ebrahimnejad2014dynamics,moller2017type}. 
Our procedure is applicable to any form of particle-boson coupling, and proceeds via efficient generation of an equation of motion (EOM) in orders of the spatial extent of bosonic clusters that arise in the dynamics. We show that this approach variationally recovers the exact infinite boson Hilbert space, provided that one converges the computation with respect to the cluster size, and we find that this is achieved with a high level of efficiency when compared against standard numerical approaches, even in the adiabatic limit. In addition to providing access to quasiparticle spectra over a wide frequency range, the GGCE comes with several strengths. In particular, it is formulated in the {\em infinite} system size limit, and thus provides access to exact spectra in the thermodynamic regime. It affords sufficient flexibility that permits extensions to finite-ranged models at finite temperatures and in higher dimensions, as well as to studies of bipolarons, and systems with different boundary conditions. Additionally, it allows the study of dynamics of non-equilibrium initial states. Lastly, since existing linear algebra solvers represent the only computational bottleneck in the approach, the GGCE serves as an easy-to-implement, methodologically unconstrained technique whose performance is limited only by access to computational resources such as large-scale parallel computing or GPU technology.

Our manuscript is organized as follows. In Section~\ref{sec:methodology}, we review the foundations of
the MA methods and devise a generalized formalism we use in the GCCE approach (Subsection~\ref{subsec:generalized AGF}). We briefly discuss our computational implementation
of the method (Subsection~\ref{subsec:algorithm}) and highlight the
relationship to and differences between our and
other methods (Subsection~\ref{subsec:comparison to other methods}). In Section~\ref{sec:results},
we demonstrate the power and scope of this implementation and present a combination of numerically exact and quasi-converged results on the Holstein~\cite{holstein1959studiesI,holstein1959studiesII}, Peierls~\cite{su1979solitons,su1980soliton,Barisic1,Barisic2} and
mixed-boson mode Holstein$+$Peierls~\cite{hannewald2004theory} models. Finally, in Section~\ref{sec:conclusion}, we conclude and discuss possible future
work.

\section{Methodology and General Considerations} \label{sec:methodology}
Consider a mobile particle (e.g. electron, hole, etc.) coupled to a bosonic field
\begin{multline} \label{field hamiltonian}
\hat H = \sum_{\veck} \varepsilon_{\veck} \hat{c}_{\veck}^\dagger \hat{c}_{\veck} + \sum_{\vecq} \hbar 
\Omega_{\vecq} \hat{b}_{\vecq}^\dagger \hat{b}_{\vecq} \\ +
\sum_{\veck,\vecq} g(\veck,\vecq) \hat{c}_{\veck+\vecq}^\dagger \hat{c}_{\veck}
(\hat{b}_{-\vecq}^\dagger + \hat{b}_\vecq).
\end{multline}
Here, the carrier (boson) has dispersion
$\varepsilon_\veck$ ($\hbar \Omega_{\vecq}$),
and the interaction $\hat{V}$ contains a vertex $g(\veck,\vecq)$
that in general depends on both
$\veck$ and $\vecq.$ We use a compact notation $\sum_\veck$ to imply a discrete sum for a problem formulated on
the lattice or a $d$-dimensional integral $\frac{l^d}{(2\pi)^d} \int \dd^d k$ with $l^d$ the system volume for a
problem in the continuum.

The goal of our approach is to derive the EOM of the one-electron Green's function at zero temperature,~\cite{mahan2013many}
\begin{equation} \label{electronic Green's function}
    G(\veck,\omega) = \mel{0}{\hat{c}_{\veck} \hat{G}(\omega)\hat{c}_{\veck}^\dagger}{0}.
\end{equation}
For Hamiltonians of the form in Eq.~\eqref{field hamiltonian}, only the retarded
component of $G(\veck, t)$ contributes,\cite{berciu2006green} and the propagator, in real frequency, takes the form
\begin{equation}
    \hat{G}(\omega) = \left[\omega - \hat{H} + \ii \eta \right]^{-1},
\end{equation}
where $\eta = 0^+$ is an artificial broadening parameter. Repeated application of Dyson's equation,
\begin{equation} \label{Dyson equation}
    \hat{G}(\omega) = \hat{G}_0(\omega) + \hat{G}(\omega) \hat{V} \hat{G}_0(\omega),
\end{equation}
with $\hat{H}_0 = \hat{H}-\hat{V}$ yields an infinite hierarchy of equations,\footnote{The connection to other approaches that utilize exact hierarchies for dynamics in polaron models (for example, the HEOM approach\cite{tanimura1989time,chen2015dynamics}) remains to be explored.} which we compute in the basis states $\ket{k, n},$ labeling a delocalized state of the carrier with definite momentum quantum number $k$ in the presence of $n$ bosons in the system. The first application of Dyson's equation yields
\begin{equation} \label{dyson identity 1}
    G(\veck, \omega) = G_0(\veck, \omega) \left[1 +
    \mel{0}{\hat{c}_\veck \hat{G}(\omega) \hat{V} \hat{c}_\veck^\dagger }{0} \right],
\end{equation}
and the second gives
\begin{equation} \label{dyson identity 2}
    \mel{0}{\hat{c}_\veck \hat{G}(\omega)\hat{V} \hat{c}_\veck^\dagger}{0} =
    \mel{0}{\hat{c}_\veck \hat{G}(\omega)\hat{V} \hat{G}_0(\omega)\hat{V} \hat{c}_\veck^\dagger}{0},
\end{equation}
where $\hat{G}_0(\omega)$ is the free particle propagator, and
\begin{equation}
    \hat{G}_0(\omega)\ket{k, n} = G_0(k, \omega - n\hbar\Omega)\ket{k, n}.
\end{equation}
Note this expansion can be indexed by the number of bosons contained
in the created states. A coupling $\hat{V}$ that is linear in boson operators either
creates or annihilates a boson, thus coupling states with $n$ bosons
to states with $n \pm 1$ bosons.

A key development 
made by Berciu~\cite{berciu2006green,berciu2007systematic} is to recast the EOM as a hierarchical
``expansion" in orders of the spatial extent of the bosonic cloud, $M$, rather than treating it as a direct expansion in the number of bosons.
Making use of the spatial structure of the
Green's functions generated through repeated application of Dyson's identity allows one to derive a scheme in which states corresponding to clouds
larger than a certain spatial extent $M$ are excluded. To illustrate the idea, consider
the example of $M = 2.$ At this level of approximation, only states with bosons localized on single and
first-neighbor sites are retained in the hierarchy. Note that this imposes no restriction on the distance between the carrier and the boson cloud. We can view this approximation as a variational ansatz in the space of Green's functions in which one allows the carrier anywhere in the system, but with bosons clustered
in a cloud of a maximum length $M.$

\begin{figure}
    \includegraphics[width=\columnwidth]{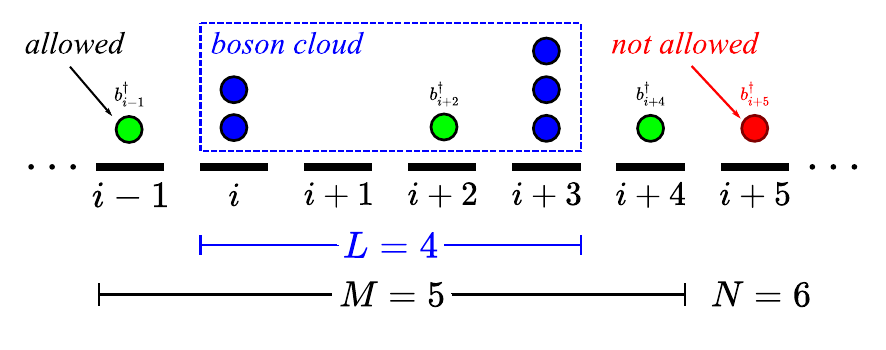}
    \caption{Cartoon of a $L=4,$ $\vecn = [2, 0, 0, 3]$ boson cloud (blue), such that $n_\mathrm{T}=5,$ contained within a variational space specified by a maximum cloud extent $M=5$ and maximum number of bosons $N=6.$ In this example, the constraint $M=5$ spans sites $i-1$ to $i+4$ so that bosons can be created only on these sites (e.g., green circles), and are not allowed outside of the $M$-site cloud (e.g., red circle). $N=6$ implies that states with two more bosons that those depicted in the figure (blue circles) are omitted from the variational space. Note that the carrier (not shown) is allowed to be anywhere on the chain.
    }
    \label{fig:cartoon1}
\end{figure}

Before delving into the details, we  provide a brief summary of the convergence parameter
space employed in our method. As discussed above, $M$ indexes the maximum extent (in units of lattice sites) of the bosonic cloud contained in the set of linear equations generated through repeated application of Dyson's
identity. Any equation in the closure must have
cloud extent $L$ such that $0 \leq L \leq M.$ The total number of bosons, $N,$ allowed in any cloud provides a second convergence parameter.  
Similarly to $M,$ only equations with a total number
of bosons $0 \leq n_\mathrm{T} \leq N$ are allowed. An
example of the configuration of a single \emph{auxiliary} Green's function
(AGF), a Green's function describing the overlap between the
single carrier and a configuration of the single carrier $+$ some boson
distribution, is given in Fig.~\ref{fig:cartoon1}. Converging $M$ and $N$ in numerical calculations allows us to approach the infinite Hilbert space limit.

Below, we detail the approach we use to construct and
solve the linear system of equations in the EOM. Specifically, in
Subsection~\ref{subsec:generalized AGF} we derive a generalized
expression for $G(k, \omega)$ for arbitrary models.
Then, in Subsection~\ref{subsec:algorithm}, we explain how
to systematically generate and solve the system of equations in computer simulations. Finally, in Subsection~\ref{subsec:comparison to other methods}, we discuss the relation of 
the GGCE to other methods. 

\subsection{A Generalized Equation of Motion}\label{subsec:generalized AGF}

We now specialize our construction to the case of one-dimensional (1D) lattice
models described by Hamiltonians of the form
\begin{equation} \label{eq: general 1d form of H}
    \hat{H} = -t \sum_{\expval{ij}} \hat{c}_i^\dagger \hat{c}_j + \Omega \sum_i \hat{b}_i^\dagger \hat{b}_i + \hat{V},
\end{equation}
where $\expval{ij}$ denotes nearest neighbors, for which numerical
results are  available, $t$ is the hopping amplitude and $\Omega$ is the frequency of dispersionless Einstein phonons. This 1D Hamiltonian allows us to both benchmark GGCE
against exact numerics and to tackle regimes that are typically difficult to study or inaccessible by related techniques even in
the well-studied 1D limit. In what follows we set $\hbar = 1$ and the lattice constant $a=1$.

Beginning with Eq.~\eqref{dyson identity 1}, we derive a generalized EOM (GEOM). Here the free particle Green's function is given by
\begin{equation}
    G_0(k, \omega) = \left[ \omega - \varepsilon_k + \ii \eta \right]^{-1},
\end{equation}
with free particle dispersion 
$\varepsilon_k = -2t \cos k$.

Consider a generalized representation of $\hat{V}$ for models that describe coupling between a carrier and a single bosonic mode,
\begin{equation} \label{generalized V_T}
    \hat V = \sum_{(g, \psi, \phi, \xi)} g \sum_i \hat{c}_i^\dagger
    \hat{c}_{i + \psi} \hat{b}_{i + \phi}^{\xi}.
\end{equation}
Here $g$ is the coupling constant,
$\psi, \phi \in \mathbb{Z}$ encode the spatial dependence of the coupling, and $\xi = \{-, +\}$ labels bosonic operators as either annihilation ($b^- \equiv b$) or creation ($b^+ \equiv b^\dagger).$ Specifically, $\psi$ is an integer that indexes the structure of the
carrier hopping in the coupling term, and $\phi$
is an integer that determines at which site relative to $i$ (the site the fermion hops to) a phonon is created.
This generalized notation completely specifies $\hat{V}$ for a given arbitary finite-ranged model. We present examples of such models in
Appendix~\ref{apdx: examples of generalized V}. For clarity, let us specialize to the Holstein model as an example:
\begin{equation} \label{Holstein V}
\begin{split}
    \hat{V}_\mathrm{H} &= \alpha \sum_i \hat{c}_i^\dagger \hat{c}_i ( \hat{b}_i^\dagger + \hat{b}_i)
\end{split}
\end{equation}
can be represented in this notation as follows
\begin{equation} \label{Holstein V2}
\begin{split}
   \hat{V}_\mathrm{H} &= \alpha \sum_i \hat{c}_i^\dagger \hat{c}_i \hat{b}_i^\dagger
   +
    \alpha \sum_i \hat{c}_i^\dagger \hat{c}_i \hat{b}_i \\
    &\leftrightarrow \{(\alpha, 0, 0, +), (\alpha, 0, 0, -)\}.
\end{split}
\end{equation}
We allow for an arbitrary but finite number of interaction terms, which need not be equal and can thus be used to model, for example,  a long-ranged coupling of a carrier to a bosonic mode.

Using Eq.~\eqref{dyson identity 1}, we arrive at the GEOM for $G(k, \omega),$
\begin{equation} \label{eq: generalized EOM}
    f_0(0) = G_0(k, \omega) \left[
    1 + \sum_{(g, \psi, \phi, \xi)} g e^{\ii kR_{\psi-\phi}}
    f_1(\phi) \right].
\end{equation}
Here, we have defined an AGF~\cite{berciu2006green,berciu2010momentum} given by
\begin{equation} \label{auxiliary f, 1d n}
    f_n(\delta) = \mathcal{N}^{-1/2}
    \sum_i e^{\ii k R_i} \mel{0}{\hat{c}_k \hat{G}(\omega) \hat{c}_{i-\delta}^\dagger
    \hat{b}_i^{\dagger n} }{0},
\end{equation}
where $\mathcal{N}$ is the number of lattice sites,
$R_m \equiv m$ and $f_n(\delta) \equiv f_n(k, \delta, \omega).$
The AGFs can be thought of as higher-order propagators of an electron in a spatial cloud composed of
multiple bosonic excitations. Further, we note the identity
$f_0(\delta) = e^{\ii k R_\delta} G(k, \omega),$ c.f.
Eq.~\eqref{auxiliary f, 1d n}.


It is now necessary to introduce additional notation for describing how AGFs with greater than zero phonons
couple. Since bosons can in general be created anywhere on
the lattice, we define an occupation number vector $\vecn,$ which labels the number of boson excitations
starting from site $i$ on a cloud embedded within the infinite lattice,
\begin{equation}
    \vecn \equiv [n^{(i)}, n^{(i+1)}, ..., n^{(i + L - 1)}],
\end{equation}
where $L \leq M$ is the length of $\vecn.$ This vector serves as a device for labeling the bosonic
Hilbert space in the following way:
$\vecn \leftrightarrow B_{i, \vecn}^\dagger \ket{0},$
where $B^\dagger_{i, \vecn} \equiv \hat{b}_i^{\dagger n_0} \hat{b}_{i+1}^{\dagger n_1} \cdots
\hat{b}_{i+L-1}^{\dagger n_{L-1}}.$ This allows us to write a generalized version of
Eq.~\eqref{auxiliary f, 1d n}, where $n$ becomes a vector,
\begin{equation} \label{auxiliary f, general}
    f_\vecn(\delta) = \mathcal{N}^{-1/2}
    \sum_i e^{\ii k R_i} \mel{0}{\hat{c}_k \hat{G}(\omega) \hat{c}_{i-\delta}^\dagger
    \hat{B}_{i, \vecn}^\dagger }{0}.
\end{equation}
Upon Fourier transforming to reciprocal space and substituting Dyson's equation, we obtain
\begin{multline}
    f_\vecn(\delta) = \mathcal{N}^{-1} \sum_{i} e^{\ii kR_i} \sum_q
    e^{-\ii q R_{i-\delta}} G_0(q, \omega - n_\mathrm{T}\Omega)\\
    \times \mel{0}{\hat{c}_k\hat{G}(\omega) \hat{V} \hat{c}_q^\dagger \hat{B}_{i, \vecn}^\dagger}{0},
\end{multline}
where $n_\mathrm{T}$ is the total number of bosons in the configuration labeled by $\vecn.$ Here we
used the fact that when
$n_\mathrm{T} > 0,$
$\mel{0}{\hat{c}_k \hat{G}_0(\omega) \hat{c}_q^\dagger \hat{B}_{i, \vecn}^\dagger}{0} = 0.$  Defining $\tilde{\omega} \equiv \omega - n_\mathrm{T}\Omega$ and adopting a combined real/momentum-space representation, we have
\begin{multline}
\label{fnEqn}
    f_\vecn(\delta) = \mathcal{N}^{-3/2} \sum_i e^{\ii k R_i} \sum_q e^{-\ii q R_{i-\delta}} G_0(q, \tilde{\omega}) \\ \times
    \sum_m e^{\ii q R_m} \mel{0}{\hat{c}_k \hat{G}(\omega)\hat{V} \hat{c}_m^\dagger
    \hat{B}_{i, \vecn}^\dagger}{0}.
\end{multline}

The goal of the procedure is to extract a relationship between AGFs with $n_\mathrm{T}$ and
$n_\mathrm{T} \pm 1$ bosons. This depends on the specific form of
$\hat{V}.$ It is thus advantageous to express $\hat{V}$ as defined in Eq.~\eqref{generalized V_T} to obtain
\begin{multline} \label{VonB}
    \hat{V} \hat{c}_m^\dagger \hat{B}_{i, \vecn}^\dagger \ket{0} = 
    \sum_{(g, \psi, \phi, \xi)} g \sum_j \hat{c}_j^\dagger
    \hat{b}_{j+\phi}^{\xi} \hat{B}_{i, \vecn}^\dagger \ket{0} \delta_{m, j+\psi}.
\end{multline}
Consider the case when $\xi = -,$ implying the boson operator removes a boson from site $j+\phi.$
Such a process can only have a non-zero contribution when a boson is removed from an occupied site, and the domain of
sites where $\hat{b}_{j+\phi}$ can act in general is $j + \phi - i \in \Gamma^-_{L} = \{0, 1, ..., L-1\}.$
In this case, an extra prefactor appears due to the boson commutation relations,
$\hat{b}_j \hat{b}_i^{\dagger m} = \hat{b}_i^{\dagger m}\hat{b}_j
+ m\delta_{ij}\hat{b}_i^{\dagger m - 1}.$

Up until now, this derivation has been exact. We now impose a limit
on the maximum cloud extent, $M,$ restricting the cluster of sites
where bosons can be created to at most $M$ connected sites, which are
occupied with up to $N$ bosons.\footnote{Note that the quasi-analytical formulation of the {\it approximate} MA methods represent a specific case of this general formalism in which $M = 1, 2$ or $3$} Thus, when $\xi = +,$ we have $j + \phi - i \in \Gamma_L^+ = \{L - M, L - M + 1, ..., M - 1\}.$ This restriction requires that we replace the sum over $j$
with a sum over the elements of the aforementioned set: 
$\sum_j \rightarrow \sum_{\gamma \in \Gamma_L^\xi}.$

To continue the derivation of the EOM, we introduce the notation: $\hat{B}_{i, \vecn}^{(\xi, \gamma)\dagger} \ket{0}$ as the state 
$\hat{B}_{i, \vecn}^\dagger \ket{0}$ with an extra boson created ($\xi = +$) or destroyed ($\xi = -$)
on site $i + \gamma$ within the permitted variational space specified by the above restriction.  We omit states indexed by $\vecn$ whose
$n_\mathrm{T} > N$ from the space of AGFs. Fig.~\ref{fig:cartoon1} demonstrates the variational space encoded in our notation.

Summing over $m$ and $q$ in Eq.~\eqref{fnEqn} produces the following general form,
\begin{multline} \label{general fn 1}
    f_\vecn(\delta) = \sum_{(g, \psi, \phi, \xi)} g \sum_{\gamma \in \Gamma_L^\xi}
    n^{(\xi, \gamma)}
    g_0(\delta + \gamma - \phi + \psi, \tilde{\omega}) \\
    \times \mathcal{N}^{-1/2} \sum_i e^{\ii k R_i}
    \mel{0}{\hat{c}_k \hat{G}(\omega) \hat{c}_{i + \gamma - \phi}^\dagger
    \hat{B}_{i, \vecn}^{(\xi, \gamma)\dagger}}{0},
\end{multline}
where $n^{(\xi, \gamma)}$ is a prefactor associated with applying a boson creation or annihilation operator: it is equal to $1$ if $\xi = +,$ and is equal to the number of bosons on site
$i+\gamma$ (before a boson is annihilated) if $\xi = -$.

Here, the free particle propagator in real space is given by~\cite{economouBook}
\begin{multline}
    g_0(\delta, \omega) = \frac{1}{\mathcal{N}}\sum_q e^{\ii q R_\delta}G_0(q, \omega) \\ 
    =  -\frac{\ii \left[-\omega_\eta/2t + \ii \sqrt{1 -(\omega_\eta/2t)^2}\right]^{|\delta|}}{ \sqrt{4t^2 - \omega_\eta^2}}, \\
    \omega_\eta \equiv \omega + \ii \eta.
\end{multline}
Observe that the second line in Eq.~\eqref{general fn 1} is precisely an AGF with different
arguments and with $n_\mathrm{T} \rightarrow n_\mathrm{T}\pm 1$ bosons. Indexing a new AGF in the same manner as before we have
\begin{multline} \label{final f general}
    f_\vecn(\delta) = \sum_{(g, \psi, \phi, \xi)} g \sum_{\gamma \in \Gamma_L^\xi}
    n^{(\xi, \gamma)} \\
    \times g_0(\delta + \gamma - \phi + \psi, \tilde{\omega})f_\vecn^{(\xi, \gamma)}(\phi - \gamma).
\end{multline}
Finally, we note that in order to abide by our labeling convention, certain
``reduction rules" for the AGFs must be followed in order to produce a valid closure. When removing
or adding bosons, as in the case $f_\vecn \rightarrow f_\vecn^{(\xi, \gamma)},$ additional phase prefactors may appear. The details of these rules are summarized in Appendix~\ref{apdx:reduction rules} (see also Ref.~\onlinecite{berciu2010momentum} for a specific example).

\subsection{Implementation} \label{subsec:algorithm}

Together, Eqs.~\eqref{eq: generalized EOM} and \eqref{final f general}, along with the rules
in Appendix~\ref{apdx:reduction rules}, contain all information necessary to solve for
$G(k, \omega)$ for some chosen values of $M, N.$ 
In this section, we describe the \emph{computational} approach for representing these equations and solving them numerically.

Every possible combination of $1 \leq n \leq N$ bosons on $1 \leq L \leq M$ sites will contribute to the calculation of $G(k, \omega).$ 
In the first step, we systematically generate all combinations,
noting the only requirement that the first
and last sites for some cloud extent $L$ must be at least singly occupied. This amounts
to symbolically constructing and storing representations of these objects, e.g.
\begin{equation}
    \mathcal{G} = \{ f_{[0]}(\delta), f_{[1]}(\delta), f_{[1, 1]}(\delta), f_{[1, 0, 2]}(\delta), ... \},
\end{equation}
such that all possible AGFs corresponding to a given configuration are
generated. This can be thought of precisely as the classic combinatorics problem of
$N$ indistinguishable balls in $M$ distinguishable bins, with the added constraint of
requiring at least one boson on each end of the cloud. In this way, the total number of
equations generated at this step (the total number of elements in $\mathcal{G},$
defined as $\abs{\mathcal{G}}$) has a straightforward representation,
\begin{equation}
    |\mathcal{G}| = 1+\sum_{L=1}^M \sum_{n=1}^N \begin{cases}
    1 & \text{if } L = 1 \text{ or } n = 2 \\
    {L + n - 3 \choose n - 2} & \text{otherwise}
    \end{cases},
\end{equation}
where the one extra term reflects the first equation in the set of equations (for $G(k,\omega)$).

The second step consists of finding the values for $\delta$ each function $f_\vecn$ requires.
Observing that the only $\delta$-dependence on the RHS of Eq.~\eqref{final f general}
is contained in $g_0$ (and importantly not in $f_\vecn$), we obtain the full closure of equations by finding, for
every $f_\vecn,$ the values of $\delta$ prescribed by the indices $\phi - \gamma$ on the RHS.
This set is informally denoted as $\mathcal{S},$ e.g.,
\begin{equation}
    \mathcal{S} = \{ f_{[0]}(-1), f_{[0]}(0), f_{[1]}(-1),... \}.
\end{equation}
The terms contained in $\mathcal{S}$ are determined by a nontrivial function of $M, N$ and depend on the
model type.  Every term in $\mathcal{S}$ is simply a specific case of the LHS of
Eq.~\eqref{final f general}. To further clarify, the
generalized equations in the set $\mathcal{G}$ leave
$\delta$ unfixed. The equations in the set $\mathcal{S}$
\emph{fix} the allowed values of $\delta$ based on the
indices $\phi - \gamma.$ We note this does not constitute an approximation to the EOM since the conditions that fix $\mathcal{S}$ arise naturally within the hierarchy.

In the final step, we formulate this as a inhomogenous linear system of equations and aim to find
the solution for all $f_\vecn(\delta)$ for some values of $k, \omega, M, N,$
\begin{equation}
    A \vecf = \vecb.
\end{equation}
Above, $A$ is a matrix of coefficients which can be read from the aforementioned equations,
and $\vecb$ is proportional to the unit vector and inherits the inhomogeneity of
Eq.~\eqref{eq: generalized EOM}.
This matrix equation can be solved in one of two ways.
The solution for $\vecf$ can be obtained in a single step, which amounts to applying some
direct solver to the $\abs{\mathcal{S}} \times \abs{\mathcal{S}}$ matrix $A.$ However, this
approach is either inefficient (using a sparse solver) or intractable using a dense solver
due to the large size of $A$ in cases such as the extreme adiabatic limit. Alternatively, we find that a continued fraction approach using dense linear
algebra provides the optimal middle ground. Formally, the
continued fractions (here we suppress the $M$ and $N$ dependence) $V_n = {\mathcal A}_n(\veck, \omega) V_{n-1} + {\mathcal B}_n(\veck, \omega) V_{n+1},$ where ${\mathcal A}_n(\veck, \omega)$ and ${\mathcal B}_n(\veck, \omega),$ are sparse matrices read off directly from the EOM, and $V_n$ is a vector of AGF's with $n\leq N$ bosons.~\cite{berciu2007systematic,berciu2010momentum} The matrix inversions required
are much smaller in this approach, although there are $\Theta(N)$ of them. We note that using this
more efficient approach, the calculations become challenging in our current implementation only around $(M, N) \sim (10, 7),$ which produces $\sim$60k equations. Adding one more boson balloons the calculation to
$\sim$150k equations, which are in principle within reach on large supercomputer architectures with sufficient memory capacity.

To approach the infinite phonon Hilbert space limit using the continued fraction approach,
we set $V_{N+1} = 0,$ solving the set of equations until we obtain $G(\veck, \omega),$ which
corresponds to $V_0.$ In the $N \rightarrow \infty$ limit, this represents a sensible boundary
condition because it becomes energetically expensive to generate clouds with larger than $N$ bosons.
In practice we treat $N$ as a convergence parameter. All results shown in this work appear to be
converged with respect to $N$ to desirable accuracy, unless otherwise stated.

\subsection{Comparison to Other Methods} \label{subsec:comparison to other methods}

\subsubsection{Comparison to related methods: Momentum Averge (MA) and Limited Phonon Basis Exact Diagonalization (LPBED) methods}
The GGCE method combines advantages from the MA and Limited Phonon Basis Exact Diagonalization\cite{de2009optical} (LPBED) methods.
In the MA approach, one makes an educated guess of the value of $M$ needed to obtain accurate
results, in essence employing a variational ansatz to the EOM. One then derives the EOM in
MA($M$) analytically ``by hand" and solves for $G(k,\omega)$ numerically. LPBED is a
more general ED analog of MA, and in principle also relies on a variational ansatz, albeit one
different from that of MA. Another successful version of LPBED\cite{de2012optical} discussed in the literature included clouds of size $M=5$ whilst allowing for two extra bosons anywhere on the lattice even away from the cloud, but with a more restricted total number of bosons.\footnote{We note that it appears the largest
values for $M$ used in the MA method is 3, while the largest values for $M$ used in the LPBED method is 5.}

We can roughly view MA and LPBED methods as specific variational
cases of the GGCE, which benefits from allowing an arbitrary
systematic choice of maximal cloud extent, $M,$ in the $N\rightarrow \infty$ limit. The GGCE thus serves as a systematically exact method which allows one to tailor resources based on the underlying physics of the problem, and is limited only by computational
resources. This provides the potential to access regimes that are difficult to quantitatively describe by other approaches, as we show below.

\subsubsection{Comparison to Variational Exact Diagonalization (VED) methods}
Variational Exact Diagonalization (VED)\cite{bonvca1999holstein}
represents another class of successful approaches to the polaron problem. In VED, a variational Hilbert space is iteratively generated by applying the off-diagonal parts of the Hamiltonian to a reference state taken to be a Bloch state of an electron and zero bosons in an infinite system. After $N_h$ iterations, one diagonalizes the Hamiltonian in the generated basis using standard Lanczos techniques. Convergence with respect to $N_h$, when possible, guarantees access to the exact ground state and a small manifold of low-lying excited states.\cite{bonca2} There are at least two main differences between GGCE and VED.  

First, VED naturally imposes a restriction on the distance between the electron and phonon configurations, which can be at most $\sim N_h$ sites (the precise value depends on the coupling), while GGCE (and MA\cite{berciu2010momentum}) includes states with the electron arbitrarily far away from the phonon clouds with no restriction (this can be seen from the application of $\hat{G}_0(\omega)$ in the EOM on states in AGFs with both an electron and phonons, which moves the electron arbitrarily in the system without regard to the location of the bosonic cloud, c.f. Eq.~\eqref{general fn 1}). We note that VED is capable of describing the ground and low-lying excited states in the weak- and lower-intermediate regimes of coupling in the adiabatic limit \cite{bonvca1999holstein}. We suspect that the restriction on the distance between the electron and the phonons in VED prohibits access to very strong couplings in the adiabatic regime and to continuum states since these are generally delocalized states (see discussion below). In contrast, as we show below, GGCE can tackle strong coupling in the adiabatic regime. 

Second, GGCE is formulated as an expansion in terms of cloud sizes, and the computation must be converged with respect to the cloud size, while VED imposes no restriction on cloud sizes (a cloud in VED can extend over, at most, $\sim N_h$ sites). For example, $N_h = 11$ implies clouds extended over $\sim$ 10-11 sites (the exact number depends on the specific model of the electron-boson coupling). Such a value of $N_h$ represents a rough lower bound within what is typically used in VED in the intermediate adiabatic limit. These values imply clouds with sizes that are much larger than those used in GGCE in the current work. This suggests that GGCE may benefit in terms of efficiency by employing a smaller number of states resulting from smaller clouds without compromising accuracy. We believe this is a direct result of using an EOM formulation of propagators, which ensures we keep only those states generated in the dynamics and nothing further. Comparing, empirically, to Ref.~\onlinecite{bonvca1999holstein}, we note that the number of states needed in GGCE appears to be two orders of magnitude smaller than those in VED in order to achieve convergence in similar parameter regimes.

Finally, we note that other variants of VED with extra restrictions on the variational space have been used with great success.\cite{bonca2,FehskeVED,ChakraVED} These, however, are either not formulated in a general enough manner to be applied to a generic form of electron-boson coupling\cite{FehskeVED} or involve further constraints that, while variational, are not completely motivated physically especially at strong couplings. In contrast, GGCE in its current form follows naturally from the EOM and has no restrictions beyond the cloud size, which is taken to the infinite limit sequentially and in an efficient manner. In principle further restrictions of this type can be imposed in our GGCE, but we do not explore this direction in the current manuscript.
 
The preliminary analysis presented here suggests that GGCE may perform more favorably than related approaches, at least in some parameter regimes and for some quantities. Future work must be devoted to address these issues and compare the range of variational restricted-basis approaches over the full range of parameter space for both ground-state energies and spectral functions to fully access the utility and efficiency of each approach.

\section{Results} \label{sec:results}
In this section, we show results for a variety of 1D lattice models described by the Hamiltonian defined
by Eqs.~\eqref{eq: general 1d form of H} and \eqref{generalized V_T}. This allows us to both benchmark GGCE
against exact numerics, and to tackle regimes typically inaccessible even in the well-studied 1D limit. In what
follows, we characterize the interaction strength via the dimensionless coupling constant
\begin{equation}
    \lambda = E_{\rm GS}(t=0)/E_{\rm GS}(\alpha=0),
\end{equation}
which is the ratio of the ground state (GS) energy in the atomic limit to that in the free
particle limit, and the adiabaticity ratio 
\begin{equation}
\Lambda = \Omega/W,
\end{equation}
where $W=4t$ is the carrier's bandwidth.

While DMC and other quantum Monte Carlo methods may access the GS in the adiabatic limit, dynamics are generally
difficult to obtain due to uncertainties associated with analytical continuation to the real-frequency axis. We showcase the ability of the GGCE to simulate dynamics in the low-energy regime for the
Holstein~\cite{holstein1959studiesI,holstein1959studiesII} (H) and Peierls (P) (also known as the
Su-Schrieffer-Heeger~\cite{su1979solitons}) models.
Finally, we study an experimentally motivated mixed Holstein$+$Peierls (HP) model in which the carrier couples to two different boson modes, one describes a Holstein coupling and the other a Peierls coupling.

\subsection{Holstein Model}

\begin{figure}[t]
    \includegraphics[width=\columnwidth]{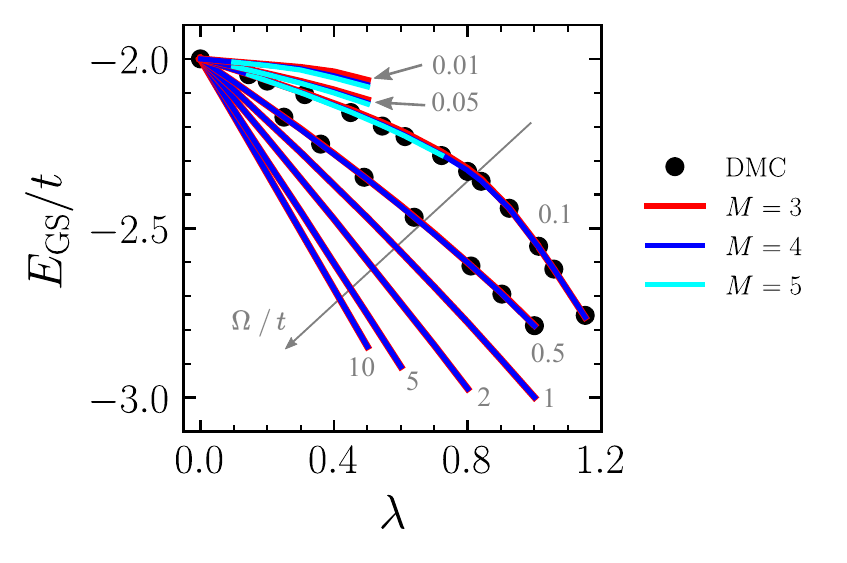}
    \caption{Ground state energy $E_\mathrm{GS}/t$ as a function of coupling strength
    $\lambda = \alpha^2/2\Omega t$ for the Holstein model in adiabaticity limits extending
    from anti- $(\Lambda \gg 1)$ to extreme-adiabatic $(\Lambda \ll 1).$
    Values for $\Omega/t$ are shown in grey, and GGCE
    results for $\Omega/t = 0.1$ and $0.5$ are compared to DMC data (symbols)
    obtained from Ref.~\onlinecite{macridin2003phonons}. Ground state peak locations are
    converged with respect to $N,$ and generally require $\approx 10$ bosons at small
    couplings, but up to $\approx 30$ at large couplings.
    }
    \label{fig:lambda-band}
\end{figure}

We first consider the prototypical Holstein model~\cite{holstein1959studiesI,holstein1959studiesII}
for which
\begin{equation}
    \hat{V}=\alpha \sum_i \hat{c}_i^\dagger \hat{c}_i (\hat{b}_i^\dagger + \hat{b}_i),
    \quad \lambda_\mathrm{H} \equiv \alpha^2/2\Omega t.
\end{equation}
In Fig.~\ref{fig:lambda-band}, we compute the GS energy of a Holstein polaron for $\Lambda \in
[0.0025,2.5].$ For $\Omega/t = 0.1$ and $0.5,$ we compare our results to those obtained via Diagrammatic Monte Carlo
(DMC).\cite{macridin2003phonons} Not only does GGCE converge to the exact result for $\lambda \in [0,1.2]$,
but it also yields slightly lower GS energies than DMC in the strong-coupling regime $\lambda  \gtrsim  1,$
although the differences are likely due to statistical errors in DMC.\footnote{We do not have access to
statistical error bars in the DMC calculations.} Importantly, we are able to converge our results to the
exact limit even at extremely small $\Omega/t \in [0.01, 0.1]$
for intermediate coupling strengths $\lambda \gtrsim 0.5,$ overcoming previous limitations of momentum average
methods. Beyond demonstrating GGCE's ability to simulate the adiabatic limit of massive bosons, our results
show a trend at intermediate couplings of the polaron binding energy $|E_{\rm GS}(\lambda) - E_{\rm GS}(0)|$
that monotonically decreases with $\Omega$.

\begin{figure*}
    \includegraphics[width=\textwidth]{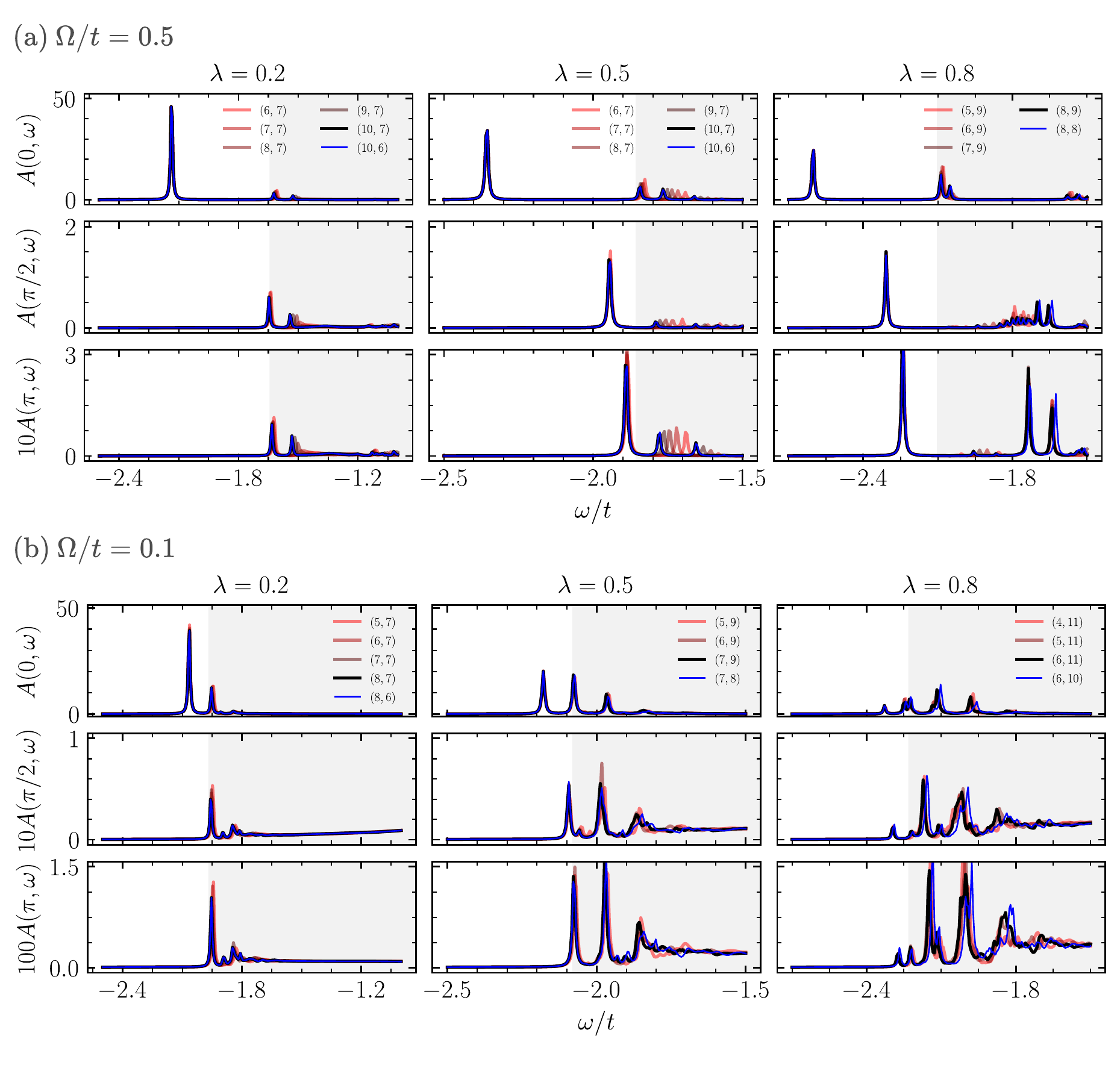}
    \centering
    \caption{Spectral function $A(k,\omega)$ at $k=0, \pi/2$ and $\pi$ 
    for the Holstein model in (a) the mildly adiatabic ($\Omega/t = 0.5$)
    and (b) adiabatic limits ($\Omega/t = 0.1$),
    for dimensionless couplings $\lambda = 0.2, 0.5$ and $0.8.$
    In this figure we use $\eta = 0.005$. Various values of $(M, N)$ are shown in the legend.
    Specifically, the gradient from red to black shows convergence with respect to $M$ for fixed
    $N.$ We demonstrate convergence with respect to $N$ via the blue line, which shows results that use
    the largest used $M,$ but with one less boson than the largest-used $N.$
    The onset of the continuum is shown in shaded gray, and is
    defined as $E_\mathrm{GS} + \Omega,$ where $E_\mathrm{GS}$ is the polaron ground-state
    energy.}
    \label{fig:HolsteinOmegaAll}
\end{figure*}

\begin{figure*}
    \centering
    \includegraphics[width=\textwidth]{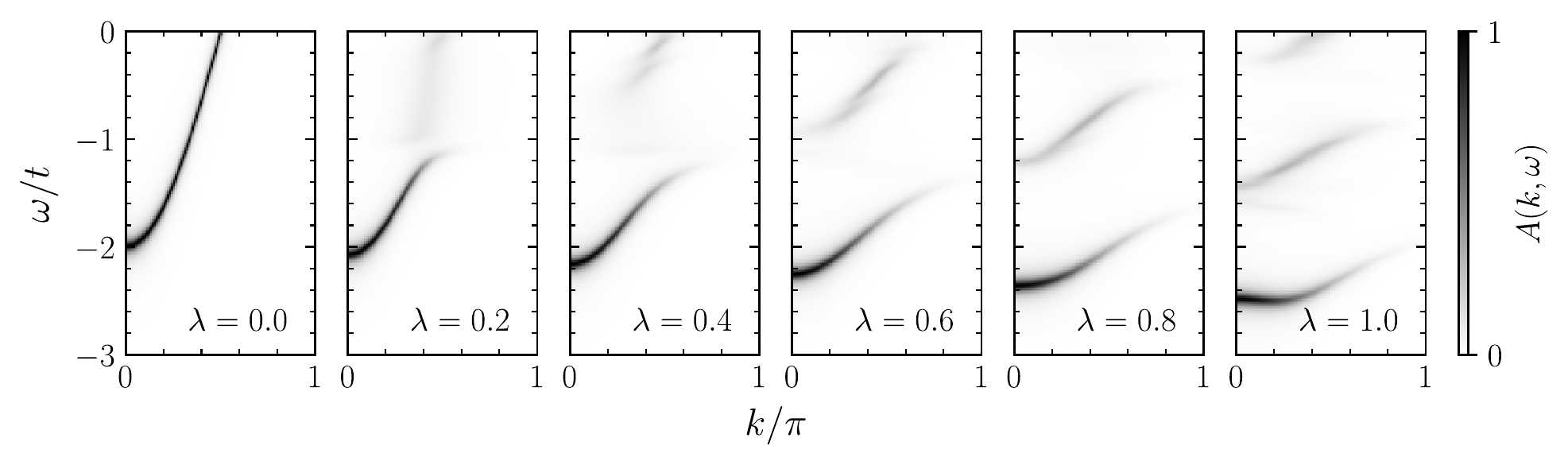}
    \caption{Spectral function $A(k,\omega)$ (scaled to a maximum of 1) of the Peierls model for
    $\Omega/t=1,$ $\eta = 0.05$ and
    various values of the dimensionless coupling strength, $\lambda.$
    $M=5$ and $N=10$ used here are sufficient for
    convergence of the bands on the scale of the plot. It is worth noting that fine structure in states above the
    lowest energy band can be resolved on a finer grid and smaller value of $\eta,$ although
    the intensity of these states is at most roughly an order of magnitude smaller than that
    of the lowest energy band.}
    \label{fig:P_bands}
\end{figure*}

To demonstrate the ability of the GGCE method to converge spectral functions and probe a broad
range of physical regimes, we present an array of spectral functions in Fig.~\ref{fig:HolsteinOmegaAll}. These results cover all combinations of $\Omega/t \in \{0.1, 0.5\},$
$k \in \{0, \pi/2, \pi\}$ and $\lambda \in \{0.2, 0.5, 0.8\}$ and highlight the potential of the method. For example, in both cases treated in Fig.~\ref{fig:HolsteinOmegaAll}, we find excellent convergence of the ground
state peak location and structure. The first excited state, which for the values of $\lambda$ considered,
lies in the polaron $+$ one boson continuum, proves more difficult to converge. Nonetheless, we show
reasonable convergence of this second peak for a wide range of parameters. However, convergence becomes more challenging 
for $\Omega/t = 0.5$ at $\lambda = 0.5,$ as seen in the second column of
Fig.~\ref{fig:HolsteinOmegaAll}(a), even when using extremely large cloud sizes ($M=10$), and as a result this peak is not sufficiently converged. Difficulty in resolving excitations above $E_\mathrm{GS} + \Omega$ is not surprising, since the nature of these
continuum states involves scattering between a delocalized electronic state and an extended cloud of phonons that is generally not small. As such, a sufficiently large cloud and therefore a bigger variational space is needed for convergence. Thus, with increasing computational resources, convergence of the spectral function proceeds naturally from low to high energy. This implies that one can readily achieve convergence of lower-energy states with much ease.

\subsection{Peierls Model}

In Fig.~\ref{fig:P_bands}, we present exact spectral functions of a polaron in the Peierls model\cite{BarisicFriedel,Barisic1,Barisic2,su1979solitons} defined by
\begin{multline}
    \hat{V}=\alpha\sum_{\expval{ij}} (\hat{c}_i^\dagger \hat{c}_j + {\rm h.c.})  (\hat{b}_i^\dagger + \hat{b}_i - \hat{b}_j^\dagger - \hat{b}_j), \\ \lambda_\mathrm{P} = 2\alpha^2/\Omega t,
\end{multline}
for a variety of different dimensionless couplings.
Although in principle no more difficult than for the case of the Holstein model, we reserve exploring the extreme-adiabatic limit ($\Omega/t \ll 1$)
to future work and show results only for $\Omega/t = 1.$

The Peierls model exhibits distinct polaron physics when compared with the Holstein model. A Peierls polaron exhibits a sharp transition from a state
with $k_{\rm GS} =0$ to one with $k_{\rm GS}\neq 0$ for $\lambda > \lambda_\mathrm{c}(\Omega/t),$ \cite{marchand2010sharp} while a Peierls bipolaron exhibits a significantly smaller mass than its Holstein counterpart\cite{SousBipolaron} and can exhibit transitions under certain conditions.\cite{SousScRep} We observe the transition to a band minimum at a finite wave vector in Fig.~\ref{fig:P_bands} as $\lambda$ changes from $\lambda = 0.8$ to $\lambda = 1,$
consistent with Ref.~\onlinecite{marchand2010sharp}. Importantly, we are able to resolve the spectrum above the ground state within sufficient accuracy. The excited states of this model play an important role in presence of other perturbations, as will become apparent next.

\subsection{Mixed-Boson Mode Holstein $+$ Peierls Model} \label{sec:mixed boson mode models}
We now consider a realistic model applicable to organic crystals, molecular complexes, etc., in which the charge carrier couples to both Holstein and Peierls phonon modes, each with its own frequency.\cite{hannewald2004theory,berkelbach2013microscopic1,berkelbach2013microscopic,fetherolf2020unification} The Hamiltonian is given by
\begin{equation} \label{H+P model}
\begin{split}
    \hat{H} =& -t \sum_{\expval{ij}} \hat{c}_i^\dagger \hat{c}_j + \Omega_{\mathrm{H}} \sum_i \hat{h}_i^{\dagger} \hat{h}_i
    + \Omega_{\mathrm{P}} \sum_i \hat{p}_i^{\dagger} \hat{p}_i  \\
    &+ \alpha_\mathrm{H} \sum_i \hat{c}_i^\dagger \hat{c}_i (\hat{h}_i^{\dagger} + \hat{h}_i) \\
    &+ \alpha_\mathrm{P} \sum_{\expval{ij}} (\hat{c}_i^\dagger \hat{c}_j + {\rm h.c.})  (\hat{p}_i^{\dagger} + \hat{p}_i - \hat{p}_j^{\dagger} - \hat{p}_j),
\end{split}
\end{equation}
where $\hat{h}_i \equiv \hat{b}_i^{\mathrm{H}},$ $\hat{p}_i \equiv \hat{b}_i^{\mathrm{P}}$
and the Holstein and Peierls boson operators act on
\emph{different} boson Hilbert spaces. We note that the combinatorics
of multi-phonon models require vastly more resources than single mode cases. Here, $\lambda_\mathrm{H} = \alpha_\mathrm{H}^2/2\Omega_\mathrm{H}t$ and $\lambda_\mathrm{P} = 2\alpha_\mathrm{P}^2/\Omega_\mathrm{P}t$, as before.

\begin{figure}
    \centering
    \includegraphics[width=\columnwidth]{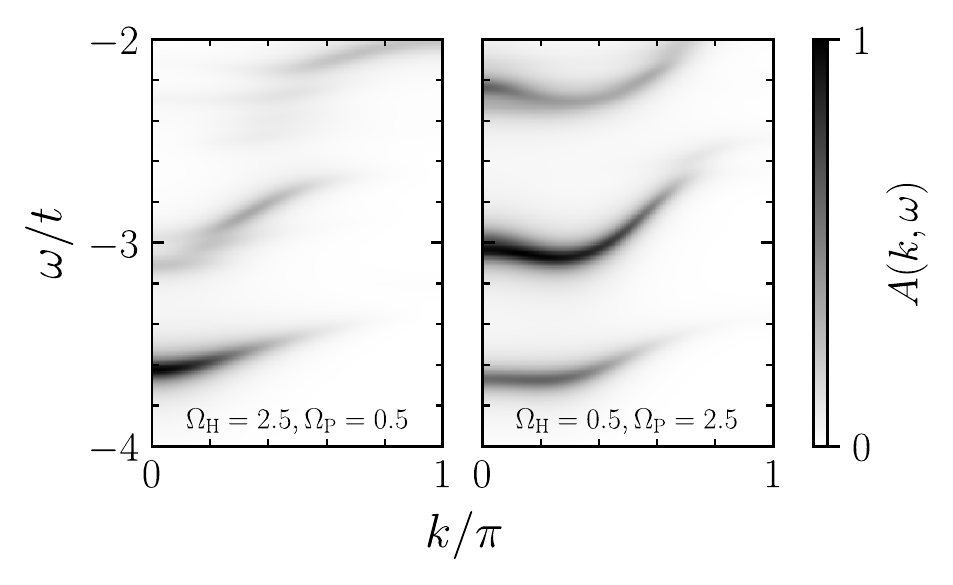}
    \caption{Spectral function $A(k,w)$ (scaled to a maximum of 1) of the mixed-boson mode Holstein$+$Peierls model for
    various values of $\Omega_\mathrm{H}$ and $\Omega_\mathrm{P},$ $\lambda_\mathrm{H}=\lambda_\mathrm{P} = 1,$ $t=1$ and
    $\eta = 0.05.$ For these calculations, we use $M_\mathrm{H} = M_\mathrm{P} = 3$, and a
    maximum total cloud length, or absolute extent, $A=3$
    (see Appendix~\ref{apdx: mixed boson mode models}), and 
    $N_\mathrm{H} = N_\mathrm{P} = 5,$  for which semi-quantitive convergence is achieved.}
    \label{fig:HP_band_1}
\end{figure}

First, we detail the differences between this HP model and that
presented in Ref.~\onlinecite{marchand2017dual}. The latter model represents
a toy model of a carrier coupled to {\em one boson type}, with two coupling contributions:
diagonal (Holstein) and off-diagonal (Peierls). Computations for this type of model possess the same scaling complexity as that for H or P models, making it much easier to converge. However, a realistic calculation requires modeling couplings to multiple phonon modes, typically of vastly different energies, characteristic of
experimental systems. A straightforward generalization of our  implementation allows us to treat the
boson modes as explicitly distinguishable, even when
$\Omega_\mathrm{P} = \Omega_\mathrm{H}$. We simply introduce two types of bosonic clouds, one for Holstein bosons with $M_\mathrm{H}$ and one for Peierls bosons with $M_\mathrm{P}$. These can overlap, and we thus need an extra variational parameter to constraint the absolute
extent, $A$, over which the combined clouds extend. We detail this construction in Appendix~\ref{apdx: mixed boson mode models}. This approach allows us to explore
this more experimentally relevant model. As previously mentioned, this
comes with the downside of increased computational
complexity. However, as we show below, we are able to semi-quantitatively converge the lowest-energy band for 
reasonably large couplings, and, with modest computational resources, we  resolve the spectrum in the
experimentally relevant regime (see Fig.~\ref{fig:HP_band_1}),
$\Omega_\mathrm{P} < \Omega_\mathrm{H},$ as well as 
in a hypothetical scenario with the frequencies reversed. This requires modest choices of $M_\mathrm{H}$, $M_\mathrm{P}$ and $A$. In Fig.~\ref{fig:HP_band_1}), in the more experimentally relevant
case, with $\Omega_\mathrm{H}/t = 2.5$ and $\Omega_\mathrm{P}/t = 0.5,$ we see a non-negligible
bandwidth and thus significant P-like character, an important observation for experiment.  In our simulations of this model, we also find interesting behavior in the second peak in the spectrum involving a minimum away from $k=0$ (not shown), which we leave to a future detailed analysis.  

\begin{figure}
    \centering
    \includegraphics[width=\columnwidth]{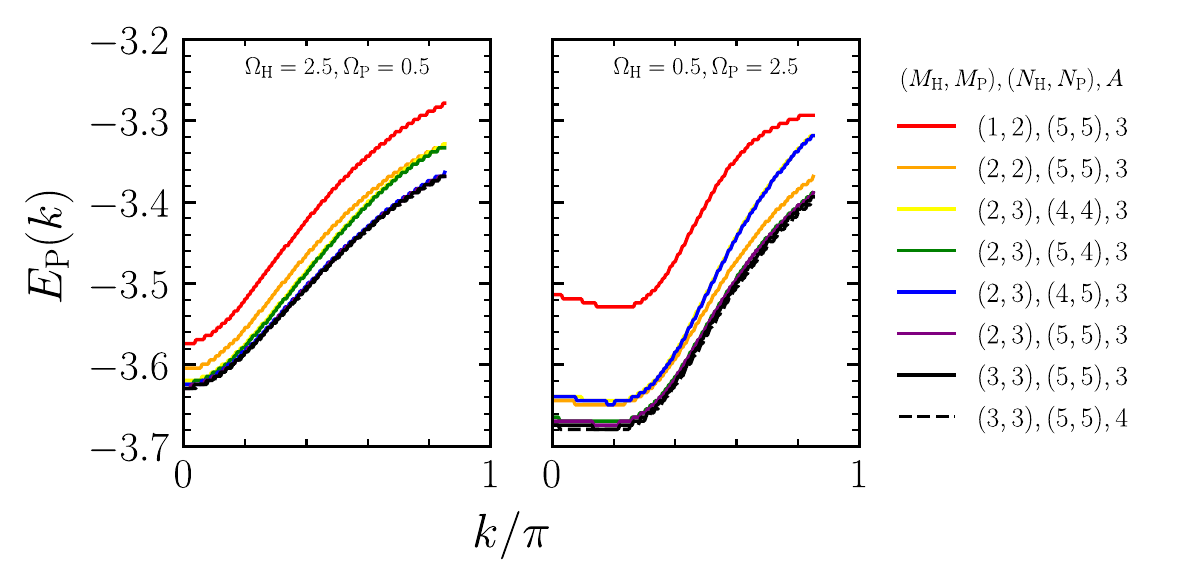}
    \caption{Convergence of the energy of the ground-state polaron band, $E_{\rm P}(k)$, for parameters shown in Fig.~\ref{fig:HP_band_1} against
    combinations of $M_\mathrm{H}, M_\mathrm{P}, A, N_\mathrm{H}$ and $N_\mathrm{P}.$}
    \label{fig:HP_gs_converge_band_1}
\end{figure}

We quantify the ground state convergence as a function of
the individual maximum cloud extents $M_\mathrm{H}, M_\mathrm{P},$ the absolute cloud extent $A$, and maximum number of bosons in the variational space,
$N_\mathrm{H}, N_\mathrm{P}$  in Fig.~\ref{fig:HP_gs_converge_band_1}. This analysis suggests that an increase of computational resources, within reach on large computers, will permit complete convergence.

\section{Conclusions} \label{sec:conclusion}
We have presented an exact, general approach to solving the EOM of a Green's function of a particle dressed by bosons, suitable for treating difficult regimes such as the adiabatic limit, and have demonstrated the power of the approach by calculating the polaron ground state and spectral functions in coupling regimes ranging from weak to strong, and adiabaticity limits ranging from extreme anti-adiabatic to extreme adiabatic. We note that at large couplings, the GGCE achieves ground state energies in agreement with DMC (Fig.~\ref{fig:lambda-band}), without the introduction of stochastic error. Exact simulated spectra for $\Lambda \ll 1$ are, in general, difficult to achieve with Monte Carlo methods due to the reliance on analytic continuation, and inaccessible to most Exact Diagonalization methods due to the large basis size needed for convergence.

We emphasize the success achieved by the MA method in characterizing
polarons and bipolarons in various systems under different physical conditions of experimental relevance.
In most of these cases, verification of the accuracy of the method against an exact approach was needed to
justify \emph{a posteriori} its utility and potential in limits where exact numerics are difficult to obtain, e.g., in higher dimensional systems. The GGCE method systematically makes use of the MA hierarchy, resulting in an exact yet efficient approach, and a new physically motivated expansion in orders of the boson cluster size, thus expanding the horizon of possibilities in characterizing dressed quasiparticles
in previously challenging regimes. Finally, the GGCE computational framework is well-suited for future practical extensions, including higher-dimensional systems, finite-temperature studies, the computation of observables connected to higher-order Green's functions, such as optical spectra and polaron mobilities,\cite{goodvin2011optical}, as well as studies of the dynamics of bipolarons,\cite{Sous2017RepBipolaron} and in other contexts we plan to address in future work.

\begin{acknowledgements}
We acknowledge helpful discussions with Kipton Barros, Robert Blackwell, Denis Golez, Kyle T. Mandli, Riccardo Rossi, David Stein, Kwangmin Yu and especially  Mona Berciu and Andrew J. Millis. We thank Alexandru Macridin for providing Quantum Monte Carlo data for the ground state energy of the Holstein polaron. M.~R.~C. acknowledges the following support: This material is based upon work supported by the U.S. Department of Energy, Office of Science, Office of Advanced Scientific Computing Research, Department of Energy Computational Science Graduate Fellowship under Award Number DE-FG02-97ER25308. Disclaimer: This report was prepared as an account of work sponsored by an agency of the United States Government. Neither the United States Government nor any agency thereof, nor any of their employees, makes any warranty, express or implied, or assumes any legal liability or responsibility for the accuracy, completeness, or usefulness of any information, apparatus, product, or process disclosed, or represents that its use would not infringe privately owned rights. Reference herein to any specific commercial product, process, or service by trade name, trademark, manufacturer, or otherwise does not necessarily constitute or imply its endorsement, recommendation, or favoring by the United States Government or any agency thereof. The views and opinions of authors expressed herein do not necessarily state or reflect those of the United States Government or any agency thereof. D.~R.~R. and J.~S. acknowledge support from the National Science Foundation (NSF) Materials Research Science and Engineering Centers (MRSEC) program through Columbia University in the Center for Precision Assembly of Superstratic and Superatomic Solids under Grant No. DMR-1420634. J.~S. also acknowledges the hospitality of the Center for Computational Quantum Physics (CCQ) at the Flatiron Institute. This research used resources of the National Energy Research Scientific Computing Center, which is supported by the Office of Science of the U.S. Department of Energy under Contract No. DE-AC02-05CH11231. We acknowledge computing resources from Columbia University's Shared Research Computing Facility project, which is supported by NIH Research Facility Improvement Grant 1G20RR030893-01, and associated funds from the New York State Empire State Development, Division of Science Technology and Innovation (NYSTAR) Contract C090171, both awarded April 15, 2010.
\end{acknowledgements}

\appendix

\section{Reduction Rules for AGFs} \label{apdx:reduction rules}
In this Appendix, we detail the reduction rules the AGFs follow in order to produce a valid EOM.

Annihilating or creating a boson to the right of the last occupied site does not come with
any additional rule for re-indexing:
\begin{multline}
    f_{[n, n', ..., n'', 0, ..., 0, 1]}(\delta) \overset{\hat{b}}{\rightarrow} f_{[n, n', ..., n'', 0, ..., 0]}(\delta) \\ = f_{[n, n', ..., n'']}(\delta),
\end{multline}
\begin{multline}
    f_{[n, n', ..., n'']}(\delta) \overset{\hat{b}^\dagger}{\rightarrow} f_{[n, n', ..., n'', 0, ..., 0, 1]}(\delta) \\ = f_{[n, n', ..., n'', 0, ..., 0, 1]}(\delta),
\end{multline}
where here $n, n'' > 0.$

However, when creating or annihilating a boson to the left of the first occupied site on the chain,
we must re-index the state such that the label $i$ always references the first occupied site:
\begin{multline}
    f_{[1, 0, ..., 0, n, n', ..., n'']}(\delta) \overset{\hat{b}}{\rightarrow} f_{[0, ..., 0, n, n', ..., n'']}(\delta) \\ \rightarrow e^{-\ii k R_z} f_{[n, n', ..., n'']}(\delta + z),
\end{multline}
\begin{multline}
    f_{[n, n', ..., n'']}(\delta) \overset{\hat{b}^\dagger}{\rightarrow} f_{[1, 0, ..., 0, n, n', ..., n'']}(\delta) \\ \rightarrow e^{\ii k R_z} f_{[1, 0, ..., 0, n, n', ..., n'']}(\delta - z),
\end{multline}
where $z$ is the number of shifted sites $i\rightarrow i\pm 1 \rightarrow i \pm 2, ...$ in the phase incurred.

\section{Examples of the Generalized Notation Used in Eq.~\eqref{generalized V_T}} \label{apdx: examples of generalized V}
In this work, we considered H, P and HP models, each of which have different
carrier-boson couplings, $\hat{V}.$ Within the framework of the GGCE, these differences amount
to a simple change in input parameters. The fully expanded coupling terms $\hat{V},$ and their
representation in terms of the notation defined in Eq.~\eqref{generalized V_T}, are
shown here. We present the three models used and reference the derivation as performed in
Section~\ref{sec:methodology}.  First, recall that the vectors which represent
the coupling are notated as $(g, \psi, \phi, \xi).$ 

In the H model, this notation translates to
\begin{equation}
    \hat{V}_\mathrm{H} = \underbrace{\alpha \sum_i \hat{c}_i^\dagger \hat{c}_i \hat{b}_i^\dagger}_{(\alpha, 0, 0, +)} +
    \underbrace{\alpha \sum_i \hat{c}_i^\dagger \hat{c}_i \hat{b}_i}_{(\alpha, 0, 0, -)}.
\end{equation}

In the P model, we have
\begin{equation}
\begin{split}
    \hat{V}_\mathrm{P} &=
    \underbrace{\alpha \sum_i \hat{c}_i^\dagger \hat{c}_{i + 1} \hat{b}_i^\dagger}_{(\alpha, 1, 0, +)}
    + \underbrace{\alpha \sum_i \hat{c}_i^\dagger \hat{c}_{i + 1} \hat{b}_i}_{(\alpha, 1, 0, -)} \\
    &\underbrace{-\alpha \sum_i \hat{c}_i^\dagger \hat{c}_{i + 1} \hat{b}_{i+1}^\dagger}_{(-\alpha, 1, 1, +)}
    \underbrace{-\alpha \sum_i \hat{c}_i^\dagger \hat{c}_{i + 1} \hat{b}_{i+1}}_{(-\alpha, 1, 1, -)} \\
    &+\underbrace{\alpha \sum_i \hat{c}_i^\dagger \hat{c}_{i - 1} \hat{b}_{i-1}^\dagger}_{(\alpha, -1, -1, +)}
    +\underbrace{\alpha \sum_i \hat{c}_i^\dagger \hat{c}_{i - 1} \hat{b}_{i-1}}_{(\alpha, -1, -1, -)} \\
    &\underbrace{-\alpha \sum_i \hat{c}_i^\dagger \hat{c}_{i - 1} \hat{b}_{i}^\dagger}_{(-\alpha, -1, 0, +)}
    \underbrace{-\alpha \sum_i \hat{c}_i^\dagger \hat{c}_{i - 1} \hat{b}_{i}}_{(-\alpha, -1, 0, -)}.
\end{split}
\end{equation}

The case of the HP model is a bit more elaborate, since the model involves different boson operators: $\hat{h}_i \equiv \hat{b}_i^{(\Omega_\mathrm{H})}$ and $\hat{p}_i \equiv \hat{b}_i^{(\Omega_\mathrm{P})}.$ Thus, we have
\begin{equation}
\begin{split}
    \hat{V}_{\mathrm{HP}} &=
    \underbrace{\alpha_\mathrm{H} \sum_i \hat{c}_i^\dagger \hat{c}_i \hat{h}_i^\dagger}_{(\alpha_\mathrm{H}, 0, 0, +)} +
    \underbrace{\alpha_\mathrm{H} \sum_i \hat{c}_i^\dagger \hat{c}_i \hat{h}_i}_{(\alpha_\mathrm{H}, 0, 0, -)} \\
    &+ \underbrace{\alpha_\mathrm{P} \sum_i \hat{c}_i^\dagger \hat{c}_{i + 1} \hat{p}_i^{\dagger}}_{(\alpha_\mathrm{P}, 1, 0, +)}
    + \underbrace{\alpha_\mathrm{P} \sum_i \hat{c}_i^\dagger \hat{c}_{i + 1} \hat{p}_i}_{(\alpha_\mathrm{P}, 1, 0, -)} \\
    &\underbrace{-\alpha_\mathrm{P} \sum_i \hat{c}_i^\dagger \hat{c}_{i + 1} \hat{p}_{i+1}^{\dagger}}_{(-\alpha_\mathrm{P}, 1, 1, +)}
    \underbrace{-\alpha_\mathrm{P} \sum_i \hat{c}_i^\dagger \hat{c}_{i + 1} \hat{p}_{i+1}}_{(-\alpha_\mathrm{P}, 1, 1, -)}\\
    &+\underbrace{\alpha_\mathrm{P} \sum_i \hat{c}_i^\dagger \hat{c}_{i - 1} \hat{p}_{i-1}^{\dagger}}_{(\alpha_\mathrm{P}, -1, -1, +)}
    +\underbrace{\alpha_\mathrm{P} \sum_i \hat{c}_i^\dagger \hat{c}_{i - 1} \hat{p}_{i-1}}_{(\alpha_\mathrm{P}, -1, -1, -)}\\
    &\underbrace{-\alpha_\mathrm{P} \sum_i \hat{c}_i^\dagger \hat{c}_{i - 1} \hat{p}_{i}^{\dagger}}_{(-\alpha_\mathrm{P}, -1, 0, +)}
    \underbrace{-\alpha_\mathrm{P} \sum_i \hat{c}_i^\dagger \hat{c}_{i - 1} \hat{p}_{i}}_{(-\alpha_\mathrm{P}, -1, 0, -)}.
\end{split}
\end{equation}

\section{Additional Notation for Mixed-Boson Mode HP Models} \label{apdx: mixed boson mode models}

In Subsection~\ref{sec:mixed boson mode models}, and specifically Fig.~\ref{fig:HP_band_1}, we introduced
new notation required to define the configuration space of the
HP model. First, the
occupation number vector $\vecn$ is now a two-row matrix,
$\overline{n},$
where as usual the columns index the site index starting with
$i,$ and the two rows correspond to the occupation numbers of
the Holstein and Peierls bosons. For clarity, we label the first
row $n_\mathrm{H}$ and the second $n_\mathrm{P}.$
The logic presented in Section~\ref{sec:methodology} still applies in for the HP model:
$\hat V$ can still create or destroy only a single boson
at a time, $\hat{B}_{i, \overline{n}}$ and corresponding
objects now reference both sets of boson occupation numbers
(and boson operators now carry a boson-type index),
$\tilde{\omega} \equiv \omega - \Omega_\mathrm{H}
\sum_j n_{\mathrm{H}}^{(j)} - \Omega_\mathrm{P} \sum_j n_{\mathrm{P}}^{(j)},$ etc. As before, the left-most occupied site is still the anchor for the
entire cloud, and thus the same reduction rules
in Appendix~\ref{apdx:reduction rules} apply.

\begin{figure}[!h]
    \includegraphics[width=\columnwidth]{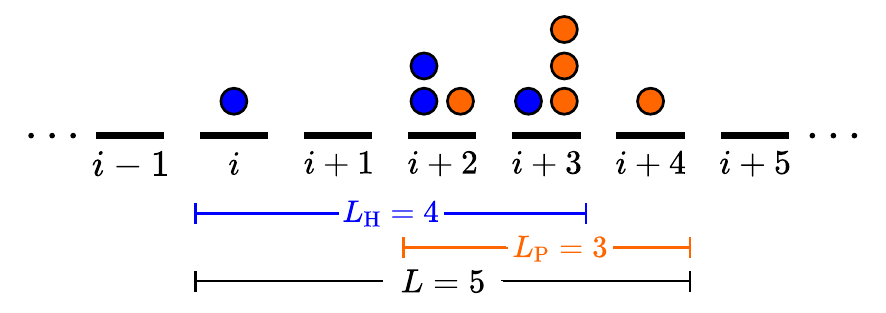}
    \caption{Example of a configuration of HP bosons corresponding to
    $n_\mathrm{H} = [1, 0, 2, 1, 0]$ and
    $n_\mathrm{P} = [0, 0, 1, 3, 1].$ Similar to the single
    boson models, we require that $\sum_j n_{\mathrm{H}}^{(j)} \leq N_\mathrm{H},$ $\sum_j n_{\mathrm{P}}^{(j)} \leq N_\mathrm{P},$
    $L_\mathrm{H} \leq M_\mathrm{H},$ $L_\mathrm{P} \leq M_\mathrm{P}$ and $L \leq A.$
    }
    \label{fig:cartoon2}
\end{figure}

In terms of the configuration space, we now limit the maximum number of Holstein and Peierls bosons individually, using
$N_\mathrm{H}$ and $N_\mathrm{P},$ respectively, and the extent of the clouds individually, using $M_\mathrm{H}$ and $M_\mathrm{P},$ respectively. Given there are now two
``overlapping" clouds of bosons which live in different
Hilbert spaces, we must define yet another configuration
space parameter, which we call the absolute extent, $A.$
This is the maximum extent of the cloud measured from the
site index of the left-most boson to the site index of the
right-most boson, regardless of boson type.
Note that we converged results in
Fig.~\ref{fig:HP_band_1} with respect to $A$ as well as the
other four convergence parameters. We present an exemplary
configuration space in Fig.~\ref{fig:cartoon2} to further
highlight the aforementioned definitions.

\newpage

\providecommand{\noopsort}[1]{}\providecommand{\singleletter}[1]{#1}%

\end{document}